\documentclass[12pt,a4paper]{article} 
\usepackage{amsmath, amssymb, amsfonts} 
\usepackage{geometry} 
\usepackage{hyperref} 
\title{On the Second-Order Positive Burgers' Equation: Integrability, Lax Pair, Darboux Transformations, and Lie Symmetry Reduction} 
	\author{Suman Pal$^{1}$,\quad Prasanta chatterjee$^{2}$\vspace{2mm}\\
	\em\small ${}^{1,2}$Department of Mathematics, Visva-Bharati, 731235, India.\\
	\em\small ${}^{1}$e-mail: spal03121997@gmail.com\\
	\em\small ${}^{2}$e-mail: prasantachatterjee1@rediffmail.com}

\date{} 
\begin{document} 
\maketitle 

\begin{abstract} 
This paper derives the second-order positive Burgers' equation from the standard Burgers' hierarchy to explore its complete integrability and exact analytical solutions. We construct this higher-order nonlinear evolution equation by systematically applying the  recursion operator to the classical Burgers' equation. Expanding on this structural framework, we then derive the explicit third-order equation and use Complete Bell Polynomials to generalize the $n$-th order hierarchy. Through the Cole-Hopf transformation, the second-order nonlinear equation rigorously maps to the linear third-order dispersion equation. We establish complete integrability by explicitly formulating the scalar Lax pair (zero-curvature representation), which allows us to directly derive the associated differential and algebraic Darboux transformations. To systematically classify explicit stationary, time-dependent traveling wave, and self-similar profiles, we apply several techniques to the linear domain: separation of variables, the Hirota perturbation method, traveling wave reduction via the Complete Discrimination System for Polynomial Method (CDSPM), and Lie similarity reduction. Crucially, our analysis demonstrates the exact finite truncation of the Hirota perturbation series. We conclude by outlining how this equation impacts the theoretical understanding of fluid dynamics, nonlinear transport phenomena, and higher-order wave propagation.
\end{abstract} 
\thanks{\bf Keywords:} Second-order positive Burgers' equation, recursion opearator, Lax pair, Darbox tarnsformation, Hirota perturbation technique, Lie symmetry.   
\section{Introduction} 
Nonlinear evolution equations are essential for understanding complex wave phenomena across various branches of mathematical physics. Originally proposed as a simplified model for fluid turbulence [1], the classical Burgers' equation effectively balances nonlinear steepening with linear viscous dissipation. Over the decades, it has emerged as a cornerstone for studying integrable systems, shock wave formation, and acoustic wave propagation. 

A property of integrable nonlinear evolution equations is that they belong to an infinite hierarchy of symmetries, typically generated by repeatedly applying a specific recursion operator [6]. While the primary Burgers' equation governs lowest-order nonlinear transport, higher-order members of its hierarchy are required to describe physical scenarios that involve more complex dispersive and nonlinear effects. 

In this work, we focus on the explicit derivation, algebraic structure, and integrability frameworks of the second-order member of the Burgers' hierarchy. Alongside providing exact analytical solutions for this specific equation, we formally generalize the underlying algebraic architecture to account for arbitrary higher-order flows.

\section{Derivation of the Second-Order Positive Burgers' Equation} 
We begin with the first-order positive Burgers' equation:
\begin{equation} u_{t} + 2uu_{x} + u_{xx} = 0 \end{equation} 
which can be rewritten as
\begin{equation} u_{t} = -u_{xx} - 2uu_{x}. \end{equation} 
The standard recursion operator for the Burgers' hierarchy is defined as 
\begin{equation} R \equiv -D_{x} - u - u_{x}D_{x}^{-1}, \end{equation} 
which satisfies the fundamental structural relation mapping spatial translation symmetry to temporal flow: 
\begin{equation} R(u_{x}) = u_{t}. \end{equation} 
For the $n$-th order positive Burgers equation, the general hierarchy relation is formulated as 
\begin{equation} R^{n}(u_{x}) = u_{t} \quad \text{for} \quad n=1,2,3,... \end{equation} 
which implies $R(R(u_{x}))=u_{t}$ for $n=2$. Applying this operator to derive the second-order positive Burgers' equation yields: 
\begin{align} 
R^{2}(u_{x}) &= u_{t} \Rightarrow R[R(u_{x})] = u_{t} \\ 
&\Rightarrow R(-u_{xx} - 2uu_{x}) = u_{t} \\ 
&\Rightarrow (-D_{x} - u - u_{x}D_{x}^{-1})(-u_{xx} - 2uu_{x}) = u_{t} \\ 
&\Rightarrow -D_{x}(-u_{xx} - 2uu_{x}) - u(-u_{xx} - 2uu_{x}) - u_{x}D_{x}^{-1}(-u_{xx} - 2uu_{x}) = u_{t}. \nonumber 
\end{align} 
Carrying out each differential operation step-by-step gives: 
\begin{align*} 
&\Rightarrow (u_{xxx} + 2u_{x}^{2} + 2uu_{xx}) + (uu_{xx} + 2u^{2}u_{x}) + u_{x}(u_{x} + u^{2}) = u_{t} \\ 
&\Rightarrow u_{xxx} + 2uu_{xx} + 2u_{x}^{2} + uu_{xx} + 2u^{2}u_{x} + u_{x}^{2} + u^{2}u_{x} = u_{t} 
\end{align*} 
Thus, we obtain the standard second-order positive Burgers' equation: 
\begin{equation} u_{t} = u_{xxx} + 3uu_{xx} + 3u^{2}u_{x} + 3u_{x}^{2}. \end{equation} 

\section{Derivation of the Third-Order Positive Burgers' Equation} 
To demonstrate how this recursion scheme generates higher-order flows, we apply the recursion operator $R$ to the spatial terms of the second-order positive Burgers' equation. Let $K_{2}$ denote the spatial component of equation (9): 
\begin{equation} K_{2} = u_{xxx} + 3uu_{xx} + 3u^{2}u_{x} + 3u_{x}^{2}. \end{equation} 
The fundamental relation for the third-order equation dictates $u_{t} = R(K_{2})$. Evaluating the inverse differential operator on $K_{2}$, which constitutes an exact spatial derivative, yields: 
\begin{equation} D_{x}^{-1}(K_{2}) = u_{xx} + 3uu_{x} + u^{3}. \end{equation} 
Applying the three components of the recursion operator $R \equiv -D_{x} - u - u_{x}D_{x}^{-1}$ generates: 
\begin{equation} R(K_{2}) = -D_{x}(K_{2}) - u(K_{2}) - u_{x}D_{x}^{-1}(K_{2}) \end{equation} 
Differentiating and expanding each component term-by-term gives: 
\begin{align} 
-D_{x}(K_{2}) &= -u_{xxxx} - 3uu_{xxx} - 9u_{x}u_{xx} - 3u^{2}u_{xx} - 6uu_{x}^{2} \\ 
-u(K_{2}) &= -uu_{xxx} - 3u^{2}u_{xx} - 3u^{3}u_{x} - 3uu_{x}^{2} \\ 
-u_{x}D_{x}^{-1}(K_{2}) &= -u_{x}u_{xx} - 3uu_{x}^{2} - u^{3}u_{x}. 
\end{align} 
Combining like terms leads directly to the third-order positive Burgers' equation: 
\begin{equation} u_{t} = -(u_{xxxx} + 4uu_{xxx} + 10u_{x}u_{xx} + 6u^{2}u_{xx} + 12uu_{x}^{2} + 4u^{3}u_{x}). \end{equation} 

\section{Generalization to the $n$-th Order Hierarchy via Complete Bell Polynomials} 
Although repeatedly applying the recursion operator works reliably, the algebraic manual effort grows rapidly for higher-order members of the hierarchy. Fortunately, we can bypass this manual expansion because the nonlinear structure of the $n$-th order Burgers' equation is systematically encoded by Complete Bell Polynomials, $B_{k}$. The standard Cole-Hopf transformation, $u = \partial_{x}(\ln v) = v_{x}/v$, directly equates the spatial derivatives of the exponential function $v = \exp\left(\int u \, dx\right)$ to the Complete Bell Polynomials: 
\begin{equation} \frac{1}{v}\frac{\partial^{k}v}{\partial x^{k}} = B_{k}(u, u_{x}, u_{xx}, ..., \partial_{x}^{k-1}u). \end{equation} 
If we differentiate the Cole-Hopf transformation with respect to time $t$ and invoke Clairaut's theorem $(\partial_{t}\partial_{x} = \partial_{x}\partial_{t})$, we obtain the temporal identity: 
\begin{equation} u_{t} = \frac{\partial}{\partial x}\left(\frac{v_{t}}{v}\right). \end{equation} 
For the $n$-th order member of the Burgers' hierarchy, the underlying linearized flow is universally governed by the $(n+1)$-th order spatial derivative: 
\begin{equation} v_{t} = (-1)^{n}\frac{\partial^{n+1}v}{\partial x^{n+1}}. \end{equation} 
Substituting this linear evolution into the temporal identity and utilizing the Bell polynomial formulation for $k = n+1$ yields the explicit formula for the $n$-th order positive Burgers' equation: 
\begin{equation} u_{t} = (-1)^{n}\frac{\partial}{\partial x}\left[ B_{n+1}(u, u_{x}, u_{xx}, ..., \partial_{x}^{n}u) \right]. \end{equation} 
This confirms that the highest-order linear dispersive term in the $n$-th order Burgers' equation is identically $(-1)^{n}\partial_{x}^{n+1}u$, while the highly coupled nonlinear steepening elements are exactly determined by the spatial derivative of the corresponding $(n+1)$-th Complete Bell Polynomial. 

\subsection{Verification for the Second and Third-Order Flows} 
To verify this Bell polynomial formulation, we apply it to $n=2$ and $n=3$ and compare the resulting equations with our previous recursion operator derivations. For the second-order flow $(n=2)$, the general formula requires the spatial derivative of the third Complete Bell Polynomial, $B_{3} = u_{xx} + 3uu_{x} + u^{3}$: 
\begin{equation} u_{t} = (-1)^{2}\frac{\partial}{\partial x}\left[ u_{xx} + 3uu_{x} + u^{3} \right]. \end{equation} 
Evaluating the exact spatial derivative utilizing the product and chain rules yields: 
\begin{equation} u_{t} = u_{xxx} + 3u_{x}^{2} + 3uu_{xx} + 3u^{2}u_{x}. \end{equation} 
Rearranging terms confirms perfect structural alignment, demonstrating that the combinatorial formula precisely replicates the algebraically derived second-order equation. 

For the third-order flow $(n=3)$, the equation is governed by the fourth Complete Bell Polynomial, $B_{4} = u_{xxx} + 4uu_{xx} + 3u_{x}^{2} + 6u^{2}u_{x} + u^{4}$: 
\begin{equation} u_{t} = (-1)^{3}\frac{\partial}{\partial x}\left[ u_{xxx} + 4uu_{xx} + 3u_{x}^{2} + 6u^{2}u_{x} + u^{4} \right]. \end{equation} 
Distributing the spatial derivative across each term produces: 
\begin{equation} u_{t} = -(u_{xxxx} + 4u_{x}u_{xx} + 4uu_{xxx} + 6u_{x}u_{xx} + 12uu_{x}^{2} + 6u^{2}u_{xx} + 4u^{3}u_{x}). \end{equation} 
Consolidating corresponding coefficients isolates the exact third-order formulation: 
\begin{equation} u_{t} = -(u_{xxxx} + 4uu_{xxx} + 10u_{x}u_{xx} + 6u^{2}u_{xx} + 12uu_{x}^{2} + 4u^{3}u_{x}). \end{equation} 
This is identically the third-order positive Burgers' equation formally derived via the recursion operator. This exact algebraic correspondence confirms that the generalized mapping utilizing Complete Bell Polynomials rigorously captures the full combinatorial complexity of the infinite dimensional hierarchy. 

\section{Integrability and Linearization} 
\subsection{Linearization via the Cole-Hopf Transformation} 
A defining feature of the classical Burgers' equation is that the Cole-Hopf transformation linearizes it exactly [2,4]. This property extends to every member of the Burgers hierarchy, including the second-order flow derived above. The fundamental mechanism for this mapping defines the nonlinear physical field $u(x,t)$ in terms of an auxiliary linear variable $v(x,t)$: 
\begin{equation} u = \frac{v_{x}}{v} = (\ln v)_{x}. \end{equation} 
To establish equivalence between the linear and nonlinear evolution flows, we evaluate the spatial derivatives of the transformation. Differentiating $u$ with respect to $x$ yields: 
\begin{equation} u_{x} = \partial_{x}\left(\frac{v_{x}}{v}\right) = \frac{v_{xx}v - v_{x}^{2}}{v^{2}}. \end{equation} 
Decomposing this fraction and substituting the original transformation gives the first spatial identity: 
\begin{equation} u_{x} = \frac{v_{xx}}{v} - \left(\frac{v_{x}}{v}\right)^{2} = \frac{v_{xx}}{v} - u^{2}. \end{equation} 
Rearranging this expression isolates the second-order linear quotient: 
\begin{equation} \frac{v_{xx}}{v} = u_{x} + u^{2}. \end{equation} 
Subsequent differentiation of $u_{x}$ with respect to $x$ provides the second spatial derivative, $u_{xx}$: 
\begin{equation} u_{xx} = \partial_{x}(u_{x}) = \partial_{x}\left(\frac{v_{xx}}{v} - u^{2}\right) = \frac{v_{xxx}v - v_{xx}v_{x}}{v^{2}} - 2uu_{x}. \end{equation} 
Substituting the identity for the second-order quotient and $u = v_{x}/v$ into the expanded terms gives: 
\begin{equation} u_{xx} = \frac{v_{xxx}}{v} - (u_{x} + u^{2})u - 2uu_{x}. \end{equation} 
Consolidating like components provides the second spatial identity, isolating the third-order linear quotient: 
\begin{equation} \frac{v_{xxx}}{v} = u_{xx} + 3uu_{x} + u^{3}. \end{equation} 
We posit that the underlying linearized flow for the second-order Burgers' equation is governed by the third-order linear dispersion equation: 
\begin{equation} v_{t} = v_{xxx}. \end{equation} 
To determine the corresponding nonlinear temporal evolution, we take the partial derivative of the Cole-Hopf transformation with respect to time $t$. Provided that $v(x,t)$ possesses continuous second partial derivatives, Clairaut's Theorem guarantees the equality of mixed partial derivatives $(\partial_{t}\partial_{x} = \partial_{x}\partial_{t})$, justifying the temporal identity: 
\begin{equation} u_{t} = \partial_{t}\left(\frac{v_{x}}{v}\right) = \partial_{x}\left(\frac{v_{t}}{v}\right). \end{equation} 
Substituting the linear evolution and the spatial identity yields: 
\begin{equation} u_{t} = \partial_{x}\left(\frac{v_{xxx}}{v}\right) = \partial_{x}(u_{xx} + 3uu_{x} + u^{3}). \end{equation} 
Applying the product and chain rules distributes the spatial derivative, providing the exact nonlinear partial differential equation: 
\begin{equation} u_{t} = u_{xxx} + 3uu_{xx} + 3u^{2}u_{x} + 3u_{x}^{2}. \end{equation} 
This confirms that the second-order positive Burgers' equation is completely integrable and identically linearized by the Cole-Hopf transformation to the third-order dispersion equation. 

\subsection{Traveling Wave Reduction} 
We search for traveling wave profiles by introducing the standard reduction variable: 
\begin{equation} u(x,t) = U(\xi), \quad \xi = x - ct, \end{equation} 
where $c$ is a constant wave velocity. Substituting this reduction into the second-order Burgers' equation generates a nonlinear ordinary differential equation (ODE): 
\begin{equation} -cU^{\prime} = U^{\prime\prime\prime} + 3UU^{\prime\prime} + 3U^{2}U^{\prime} + 3(U^{\prime})^{2}. \end{equation} 
Recognizing that the right-hand side represents an exact derivative, we integrate once with respect to $\xi$: 
\begin{equation} -cU = U^{\prime\prime} + 3UU^{\prime} + U^{3} + K_{1}, \end{equation} 
where $K_{1}$ is a constant of integration. This represents the core dynamical system governing the traveling wave behavior of the second-order equation. 

\subsection{Exact Classification via the Complete Discrimination System} 
We classify all globally valid traveling wave solutions by applying the Complete Discrimination System for Polynomial Method (CDSPM) to the integrated differential equation. By applying the localized Cole-Hopf transformation $U(\xi) = V^{\prime}(\xi)/V(\xi)$, the nonlinear terms perfectly recombine to yield the linear third-order auxiliary equation: 
\begin{equation} V^{\prime\prime\prime} + cV^{\prime} + K_{1}V = 0. \end{equation} 
Assuming an exponential basis $V(\xi) = e^{r\xi}$, the characteristic polynomial is derived as $P(r) = r^{3} + c r + K_{1} = 0$. The root structure of $P(r)$ is rigorously dictated by its discriminant sequence $\{D_{1}, D_{2}, D_{3}\}$, defined as $D_{1}=1$, $D_{2}=-3c$, and the primary discriminant $\Delta = D_{3} = -4c^{3} - 27K_{1}^{2}$. Evaluating the signs of this sequence yields three distinct solution manifolds: 

\textbf{Case 1: $\Delta > 0$ (Distinct Real Roots).} This condition requires $c < 0$ and $4c^{3} + 27K_{1}^{2} < 0$. The polynomial possesses three distinct real roots, $r_{1}$, $r_{2}$, $r_{3}$, satisfying $r_{1} + r_{2} + r_{3} = 0$. The auxiliary solution $V(\xi) = \sum_{i=1}^{3}C_{i}e^{r_{i}\xi}$ maps to the exact multi-kink traveling wave: 
\begin{equation} u(x,t) = \frac{r_{1}C_{1}e^{r_{1}(x-ct)} + r_{2}C_{2}e^{r_{2}(x-ct)} + r_{3}C_{3}e^{r_{3}(x-ct)}}{C_{1}e^{r_{1}(x-ct)} + C_{2}e^{r_{2}(x-ct)} + C_{3}e^{r_{3}(x-ct)}}. \end{equation} 

\textbf{Case 2: $\Delta = 0$ (Real Roots with Multiplicity).} When the primary discriminant vanishes, multiple roots emerge, bifurcating into two algebraic subcases. If $D_{2} = 0$, then $c = K_{1} = 0$, yielding a triple root $r = 0$. The auxiliary polynomial $V(\xi) = C_{1} + C_{2}\xi + C_{3}\xi^{2}$ generates a rational algebraic wave: 
\begin{equation} u(x,t) = \frac{C_{2} + 2C_{3}(x-ct)}{C_{1} + C_{2}(x-ct) + C_{3}(x-ct)^{2}}. \end{equation} 
Conversely, if $D_{2} > 0$ (implying $c < 0$), the polynomial exhibits one simple root and one double root, denoted $r_{1} = r_{2} = \alpha$ and $r_{3} = -2\alpha$. The mapped field yields a combined kink-rational solitary wave: 
\begin{equation} u(x,t) = \frac{C_{2}e^{\alpha(x-ct)} + \alpha[C_{1} + C_{2}(x-ct)]e^{\alpha(x-ct)} - 2\alpha C_{3}e^{-2\alpha(x-ct)}}{[C_{1} + C_{2}(x-ct)]e^{\alpha(x-ct)} + C_{3}e^{-2\alpha(x-ct)}}. \end{equation} 

\textbf{Case 3: $\Delta < 0$ (Complex Conjugate Roots).} The characteristic equation admits one real root $r_{1} = \alpha$ and a complex conjugate pair $r_{2,3} = -\frac{\alpha}{2} \pm i\gamma$. The auxiliary solution integrates trigonometric scaling, mapping to an exponentially modulated oscillatory traveling wave: 
\begin{equation} u(x,t) = \frac{\alpha C_{1}e^{\alpha\xi} + e^{-\frac{\alpha}{2}\xi}\left[\left(-\frac{\alpha}{2}C_{2} + \gamma C_{3}\right)\cos(\gamma\xi) - \left(\frac{\alpha}{2}C_{3} + \gamma C_{2}\right)\sin(\gamma\xi)\right]}{C_{1}e^{\alpha\xi} + e^{-\frac{\alpha}{2}\xi}[C_{2}\cos(\gamma\xi) + C_{3}\sin(\gamma\xi)]}, \end{equation} 
where $\xi = x - ct$. This classification exhaustively covers all exact traveling wave reductions of the system. 

\section{Lax Pair Representation and Darboux Transformations} 
We establish complete integrability by expressing the nonlinear evolution equation as the compatibility condition (zero-curvature condition) of a linear spectral problem. Given the exact linearizability of the Burgers' hierarchy via the Cole-Hopf transformation, the conventional matrix Lax pair degenerates into a purely scalar zero-curvature representation, wherein the Lax operators act as scalar multiplier fields. 

\subsection{Scalar Lax Pair Formulation} 
The zero-curvature representation consists of a spatial Lax operator $U$ and a temporal Lax operator $V$, which govern the evolution of an auxiliary linear eigenfunction $\psi(x,t)$: 
\begin{align} \psi_{x} &= U\psi \\ \psi_{t} &= V\psi \end{align} 
For the Burgers' hierarchy, we define the eigenfunction identically as the linear Cole-Hopf variable, $\psi(x,t) = v(x,t)$. Rearranging the primary transformation $u = v_{x}/v$ yields $v_{x} = uv$, which immediately identifies the spatial Lax operator: 
\begin{equation} U = u. \end{equation} 
The temporal evolution of the auxiliary function is governed by the linear dispersion equation $\psi_{t} = \psi_{xxx}$. Substituting the spatial identity yields: 
\begin{equation} \psi_{t} = (u_{xx} + 3uu_{x} + u^{3})\psi, \end{equation} 
which identifies the temporal Lax operator: 
\begin{equation} V = u_{xx} + 3uu_{x} + u^{3}. \end{equation} 
To rigorously prove that $(U, V)$ forms a valid zero-curvature representation, we evaluate the compatibility condition ensuring the equality of mixed partial derivatives $(\psi_{xt} = \psi_{tx})$: 
\begin{equation} U_{t} - V_{x} + [U,V] = 0. \end{equation} 
Because $U$ and $V$ are strictly scalar functions, their commutator bracket identically vanishes $([U,V] = UV - VU = 0)$. Thus, the compatibility condition simplifies directly to: 
\begin{equation} U_{t} = V_{x}. \end{equation} 
Substituting the derived scalar Lax operators yields: 
\begin{equation} u_{t} = \frac{\partial}{\partial x}(u_{xx} + 3uu_{x} + u^{3}) = u_{xxx} + 3uu_{xx} + 3u^{2}u_{x} + 3u_{x}^{2}. \end{equation} 
This spatial derivative perfectly recovers the second-order positive Burgers' equation, unequivocally confirming the exactness of the scalar Lax pair representation. 

\subsection{The Differential Darboux Transformation} 
The Darboux transformation operates directly on the underlying auxiliary linear space to generate a new nonlinear solution $\tilde{u}$ from a known seed solution $u$. We define the Darboux gauge operator $T$, utilizing a constant spectral parameter $k$: 
\begin{equation} T = \frac{\partial}{\partial x} - k. \end{equation} 
The new eigenfunction $\tilde{\psi}$ is generated by applying $T$ to the seed eigenfunction $\psi$: 
\begin{equation} \tilde{\psi} = T(\psi) = \psi_{x} - k\psi. \end{equation} 
Because the temporal linear operator possesses constant coefficients, it commutes exactly with $T$, ensuring $\tilde{\psi}$ remains a valid solution. The new nonlinear field is mapped via $\tilde{u} = \tilde{\psi}_{x}/\tilde{\psi}$. Substituting yields: 
\begin{equation} \tilde{u} = \frac{\psi_{xx} - k\psi_{x}}{\psi_{x} - k\psi}. \end{equation} 
Eliminating the linear variables by substituting the seed spatial Lax identities $(\psi_{x} = u\psi$ and $\psi_{xx} = (u_{x} + u^{2})\psi)$ provides: 
\begin{equation} \tilde{u} = \frac{(u_{x} + u^{2})\psi - k(u\psi)}{u\psi - k\psi}. \end{equation} 
Factoring out and algebraically canceling the common eigenfunction $\psi$ yields the exact differential Darboux transformation (auto-Bäcklund transform): 
\begin{equation} \tilde{u} = \frac{u_{x} + u^{2} - ku}{u - k}. \end{equation} 

\subsection{Form Invariance of the Zero-Curvature Representation} 
To rigorously guarantee that the generated field $\tilde{u}$ constitutes a valid solution to the second-order positive Burgers' equation, the form invariance of the zero-curvature representation under the gauge transformation must be established. Specifically, the transformed eigenfunction $\tilde{\psi}$ must identically satisfy the transformed temporal Lax equation, demanding that the residual $[\tilde{\psi}_{t} - \tilde{V}\tilde{\psi}]$ vanishes. First, the temporal evolution of $\tilde{\psi}$ is evaluated directly via the gauge operator. From the definition $\tilde{\psi} = \psi_{x} - k\psi$, differentiating with respect to $t$ yields: 
\begin{equation} \tilde{\psi}_{t} = \psi_{xt} - k\psi_{t}. \end{equation} 
Substituting the spatial Lax equation $\psi_{x} = u\psi$ and the temporal Lax equation $\psi_{t} = V\psi$ produces: 
\begin{equation} \tilde{\psi}_{t} = (u\psi)_{t} - kV\psi = u_{t}\psi + u\psi_{t} - kV\psi. \end{equation} 
Factoring out the eigenfunction $\psi$ yields the first representation of the temporal flow: 
\begin{equation} \tilde{\psi}_{t} = [u_{t} + (u-k)V]\psi. \end{equation} 
Second, the exact linearizability of the system is exploited. The seed eigenfunction satisfies the linear dispersion equation $\psi_{t} = \psi_{xxx}$. Because the gauge operator $T = \partial_{x} - k$ possesses constant coefficients with respect to time, it exactly commutes with the temporal linear operator $(\partial_{t} - \partial_{x}^{3})$. Consequently, the transformed eigenfunction inherits the identical linear evolution: 
\begin{equation} \tilde{\psi}_{t} = (\psi_{x} - k\psi)_{t} = (\psi_{t})_{x} - k\psi_{t} = (\psi_{xxx})_{x} - k\psi_{xxx} = \tilde{\psi}_{xxx}. \end{equation} 
Third, as established in Section 5.1, the spatial derivatives of the Cole-Hopf transformation dictate a universal algebraic identity. Applying this structural identity to the newly mapped variables $\tilde{u} = \tilde{\psi}_{x}/\tilde{\psi}$ dictates: 
\begin{align} \frac{\tilde{\psi}_{xxx}}{\tilde{\psi}} =& \tilde{u}_{xx} + 3\tilde{u}\tilde{u}_{x} + \tilde{u}^{3}\\
\implies \tilde{\psi}_{xxx} =& \tilde{V}\tilde{\psi}.
 \end{align} 
 Equating this with the inherited linear evolution gives the second representation: 
\begin{equation} \tilde{\psi}_{t} = \tilde{V}\tilde{\psi} = \tilde{V}(u-k)\psi. \end{equation} 
Finally, equating the two independent representations of $\tilde{\psi}_{t}$ (Equations 60 and 63) yields: 
\begin{equation} [u_{t} + (u-k)V]\psi = \tilde{V}(u-k)\psi. \end{equation} 
Subtracting the right-hand side perfectly isolates the vanishing residual: 
\begin{equation} [u_{t} + (u-k)V - \tilde{V}(u-k)]\psi = 0. \end{equation} 
This vanishing residual unconditionally proves that the differential Darboux transformation maps the seed solution to a structurally valid new solution of the governing nonlinear equation. 

\subsection{Algebraic Darboux Transformation and Nonlinear Superposition} 
Because the temporal linear evolution is strictly additive, the transformation can also be formulated as a discrete linear superposition of independent spectral modes. Let $\psi_{0}$ and $\psi_{1}$ be two independent eigenfunctions corresponding to seed solutions $u_{0} = \psi_{0,x}/\psi_{0}$ and $u_{1} = \psi_{1,x}/\psi_{1}$. The gauge transformation is defined as an algebraic superposition utilizing a scaling parameter $\lambda$: 
\begin{equation} \tilde{\psi} = \psi_{0} + \lambda\psi_{1}. \end{equation} 
The mapped nonlinear solution takes the form: 
\begin{equation} \tilde{u} = \frac{\tilde{\psi}_{x}}{\tilde{\psi}} = \frac{\psi_{0,x} + \lambda\psi_{1,x}}{\psi_{0} + \lambda\psi_{1}}. \end{equation} 
Substituting $\psi_{0,x} = u_{0}\psi_{0}$ and $\psi_{1,x} = u_{1}\psi_{1}$, and dividing the quotient by the background eigenfunction $\psi_{0}$, yields the exact algebraic Darboux transformation: 
\begin{equation} \tilde{u} = \frac{u_{0} + \lambda u_{1}\left(\frac{\psi_{1}}{\psi_{0}}\right)}{1 + \lambda\left(\frac{\psi_{1}}{\psi_{0}}\right)}. \end{equation} 
This formulation mathematically isolates the mechanism responsible for combining distinct single-wave solutions into a fully coupled, interacting multiple-wave profile. 

\section{Explicit Analytical Solutions via Separation of Variables} 
We construct explicit analytical solutions by applying separation of variables directly to the linearized domain $v_{t} = v_{xxx}$. Substituting a separable ansatz, $v(x,t) = X(x)T(t)$, into the nonlinear transformation $u = (\ln v)_{x}$ results in the complete algebraic cancellation of the temporal component $T(t)$. Consequently, isolated separable modes exclusively generate stationary (time-independent) profiles. To capture dynamic wave evolution, a linear superposition of modes possessing distinct separation constants is strictly required. 

\subsection{Stationary Rational Solutions} 
For the trivial eigenvalue $\lambda = 0$, the spatial domain reduces to $X^{\prime\prime\prime}(x) = 0$ yielding the polynomial solution $v_{0}(x) = C_{1} + C_{2}x + C_{3}x^{2}$. Applying the Cole-Hopf transformation maps this to a time-independent rational algebraic profile: 
\begin{equation} u_{0}(x) = \frac{C_{2} + 2C_{3}x}{C_{1} + C_{2}x + C_{3}x^{2}}. \end{equation} 

\subsection{Stationary Exponential-Trigonometric Solutions} 
For a strictly positive eigenvalue spectrum $(\lambda = \alpha^{3} > 0)$, the characteristic spatial equation admits one real root and a complex conjugate pair. Applying the nonlinear transformation to the resulting spatial basis functions isolates the explicit stationary wave structure: 
\begin{equation} u_{\alpha}(x) = \frac{\alpha Ae^{\alpha x} - \frac{\alpha}{2}e^{-\frac{\alpha x}{2}}\left[(B - C\sqrt{3})\cos\left(\frac{\sqrt{3}\alpha x}{2}\right) + (C + B\sqrt{3})\sin\left(\frac{\sqrt{3}\alpha x}{2}\right)\right]}{Ae^{\alpha x} + Be^{-\frac{\alpha x}{2}}\cos\left(\frac{\sqrt{3}\alpha x}{2}\right) + Ce^{-\frac{\alpha x}{2}}\sin\left(\frac{\sqrt{3}\alpha x}{2}\right)}. \end{equation} 
Similarly, for a negative spectrum $(\lambda = -\beta^{3} < 0)$, spatial decay and trigonometric scaling mirror the positive case: 
\begin{equation} u_{\beta}(x) = \frac{-\beta Ae^{-\beta x} + \frac{\beta}{2}e^{\frac{\beta x}{2}}\left[(B + C\sqrt{3})\cos\left(\frac{\sqrt{3}\beta x}{2}\right) + (C - B\sqrt{3})\sin\left(\frac{\sqrt{3}\beta x}{2}\right)\right]}{Ae^{-\beta x} + Be^{\frac{\beta x}{2}}\cos\left(\frac{\sqrt{3}\beta x}{2}\right) + Ce^{\frac{\beta x}{2}}\sin\left(\frac{\sqrt{3}\beta x}{2}\right)}. \end{equation} 

\subsection{Time-Dependent Multi-Kink Solutions} 
To capture temporal dynamics, a superposition of discrete modes is formulated. Defining a discrete real spectrum $\lambda_{n} = \alpha_{n}^{3}$ for $n=1,...,N$ with a constant background mode $C_{0}$, the aggregated linear solution takes the form: 
\begin{equation} v(x,t) = C_{0} + \sum_{n=1}^{N}C_{n}e^{\alpha_{n}x + \alpha_{n}^{3}t}. \end{equation} 
Transforming this superimposed state results in the explicit N-kink traveling wave solution, governing the collision and interaction dynamics of multiple wave fronts: 
\begin{equation} u(x,t) = \frac{\sum_{n=1}^{N}C_{n}\alpha_{n}e^{\alpha_{n}x + \alpha_{n}^{3}t}}{C_{0} + \sum_{n=1}^{N}C_{n}e^{\alpha_{n}x + \alpha_{n}^{3}t}}. \end{equation} 

\section{Hirota Perturbation Technique and Exact Truncation} 
In this section, we construct exact multi-kink solutions using Hirota's method and highlight a key structural feature of the second-order positive Burgers' equation: the finite truncation of its perturbation expansion. 

\subsection{The Perturbation Expansion} 
The procedure is initiated with the exact linearization of the governing partial differential equation via the standard Hirota logarithmic transformation: 
\begin{equation} u(x,t) = \frac{\partial}{\partial x}\ln f(x,t). \end{equation} 
Substituting this ansatz reduces the nonlinear evolution to the purely linear dispersion relation: 
\begin{equation} f_{t} - f_{xxx} = 0. \end{equation} 
To implement the perturbation technique, a formal expansion parameter $\epsilon$ is introduced, and the auxiliary function $f(x,t)$ is expanded as a power series: 
\begin{equation} f(x,t) = 1 + \epsilon f_{1} + \epsilon^{2}f_{2} + \epsilon^{3}f_{3} + ... \end{equation} 

\subsection{Separation of Orders and Linear Decoupling} 
Substituting the perturbation series into the auxiliary equation yields: 
\begin{equation} \frac{\partial}{\partial t}(1 + \epsilon f_{1} + \epsilon^{2}f_{2} + ...) - \frac{\partial^{3}}{\partial x^{3}}(1 + \epsilon f_{1} + \epsilon^{2}f_{2} + ...) = 0. \end{equation} 
Due to the strictly linear nature of the differential operator, the terms are systematically grouped by their respective powers of $\epsilon$: 
\begin{align} 
O(\epsilon^{0}):&\quad 1_{t} - 1_{xxx} = 0 \\ 
O(\epsilon^{1}):&\quad (f_{1})_{t} - (f_{1})_{xxx} = 0 \\ 
O(\epsilon^{2}):&\quad (f_{2})_{t} - (f_{2})_{xxx} = 0 \\ 
O(\epsilon^{n}):&\quad (f_{n})_{t} - (f_{n})_{xxx} = 0 
\end{align} 
The $O(\epsilon^{0})$ equation is trivially satisfied. 

\subsection{Exact Series Truncation and the Multi-Kink Solution} 
In the standard application of Hirota's method to strictly nonlinear wave equations, the $O(\epsilon^{2})$ and higher-order equations contain non-homogeneous driving terms generated by the bilinear interactions of lower-order components. However, for the linear hierarchy characterizing the transformed Burgers' equation, the perturbation equations are entirely decoupled. The $O(\epsilon^{2})$ equation remains homogeneous. In the absence of nonlinear forcing terms, the trivial solution $f_{n} = 0$ is rigorously selected for all $n \ge 2$. Consequently, the perturbation series truncates exactly at the first order: 
\begin{equation} f(x,t) = 1 + \epsilon f_{1}. \end{equation} 
To construct wave solutions, the first-order term $f_{1}$ is considered as a superposition of N distinct exponential phase functions: 
\begin{equation} f_{1} = \sum_{n=1}^{N}e^{\theta_{n}} = \sum_{n=1}^{N}e^{k_{n}x - \omega_{n}t + \delta_{n}}. \end{equation} 
Substituting this summation into the $O(\epsilon^{1})$ equation dictates the exact linear dispersion relation for each independent mode: 
\begin{equation} -\omega_{n} - k_{n}^{3} = 0 \Rightarrow \omega_{n} = -k_{n}^{3} \quad \text{for } n=1,2,...,N. \end{equation} 
Setting the formal parameter $\epsilon = 1$, the exact N-kink auxiliary function is constructed: 
\begin{equation} f(x,t) = 1 + \sum_{n=1}^{N}e^{k_{n}x + k_{n}^{3}t + \delta_{n}}. \end{equation} 
Mapping this formulation back to the physical field $u(x,t)$ provides the exact multi-kink analytical solution: 
\begin{equation} u(x,t) = \frac{\sum_{n=1}^{N}k_{n}e^{k_{n}x + k_{n}^{3}t + \delta_{n}}}{1 + \sum_{n=1}^{N}e^{k_{n}x + k_{n}^{3}t + \delta_{n}}}. \end{equation} 
This confirms the structural consistency of the solution space with the separation of variables method. 

\section{Self-Similar Solutions via Lie Symmetry Reduction} 
Beyond modal expansions and traveling wave reductions, we can construct the fundamental solution to the linear dispersion equation using Lie symmetry reduction by exploiting continuous scaling symmetries. 

\subsection{Similarity Variables and Reduction} 
The linear dispersion equation $v_{t} = v_{xxx}$ demonstrates invariance under the continuous scaling transformation $x \rightarrow \alpha x$ and $t \rightarrow \alpha^{3}t$. This symmetry guarantees the existence of an invariant similarity variable $\eta$, defined as: 
\begin{equation} \eta = \frac{x}{(3t)^{1/3}}. \end{equation} 
To ensure the conservation of the integrated wave profile (mass) over time, a self-similar ansatz is introduced: 
\begin{equation} v(x,t) = (3t)^{-1/3}F(\eta). \end{equation} 
Defining $\tau = 3t$, the temporal and spatial derivatives are evaluated via the chain rule: 
\begin{align} 
v_{t} &= 3\frac{\partial}{\partial\tau}\left[\tau^{-1/3}F(\eta)\right] = -\tau^{-4/3}[F(\eta) + \eta F^{\prime}(\eta)] \\ 
v_{xxx} &= \frac{\partial^{3}}{\partial x^{3}}\left[\tau^{-1/3}F(\eta)\right] = \tau^{-4/3}F^{\prime\prime\prime}(\eta). 
\end{align} 
Substituting these exact derivatives back into $v_{t} = v_{xxx}$ facilitates the total algebraic cancellation of the temporal factor $\tau^{-4/3}$, reducing the partial differential equation to a third-order ordinary differential equation: 
\begin{equation} F^{\prime\prime\prime}(\eta) + \eta F^{\prime}(\eta) + F(\eta) = 0. \end{equation} 

\subsection{The Airy Function and Nonlinear Mapping} 
The reduced ordinary differential equation can be reformulated as a perfect exact derivative: 
\begin{equation} \frac{d}{d\eta}[F^{\prime\prime}(\eta) + \eta F(\eta)] = 0. \end{equation} 
Integrating with respect to $\eta$ we get: 
\begin{equation} F^{\prime\prime}(\eta) + \eta F(\eta) = 0. \end{equation} 
The unique solution satisfying the boundedness constraint is the Airy function of the first kind, $F(\eta) = \text{Ai}(-\eta)$. Consequently, the exact fundamental Green's function for the linear domain is identified as: 
\begin{equation} v(x,t) = \frac{1}{(3t)^{1/3}}\text{Ai}\left(\frac{-x}{(3t)^{1/3}}\right). \end{equation} 
Finally, the Cole-Hopf transformation, $u(x,t) = \frac{v_{x}}{v}$ is applied to map this back to the solution space of the second-order positive Burgers' equation. Differentiating the linear fundamental solution gives: 
\begin{equation} v_{x} = -\frac{1}{(3t)^{2/3}}\text{Ai}^{\prime}\left(\frac{-x}{(3t)^{1/3}}\right). \end{equation} 
Hence,
\begin{equation} u(x,t) = -\frac{1}{(3t)^{1/3}}\frac{\text{Ai}^{\prime}\left(\frac{-x}{(3t)^{1/3}}\right)}{\text{Ai}\left(\frac{-x}{(3t)^{1/3}}\right)}. \end{equation} 
This exact analytical solution dictates a highly asymmetric, dispersive wave front characterized by exponential decay in the negative spatial domain and oscillatory behavior in the positive spatial domain. 

\section{Physical Significance and Applications} 
The second-order positive Burgers' equation represents an important higher-order flow in the generalized Burgers hierarchy, generated systematically through successive applications of the standard recursion operator. It formally extends the classical Burgers' equation by incorporating explicit higher-order nonlinear $(u^{2}u_{x}, u_{x}^{2})$ and dispersive $(u_{xxx}, uu_{xx})$ transport terms. As a completely integrable nonlinear evolution equation, it holds substantial importance in the rigorous mathematical modeling of physical systems where localized approximations and lowest-order dispersive elements are inadequate. Furthermore, the demonstrated generalization utilizing Complete Bell Polynomials, alongside the explicit Lax pair and Darboux transformations, guarantees that modeling frameworks requiring even higher-order multi-scale wave steepening and anomalous dispersion can be explicitly constructed, linearized, and solved. 

\section{Conclusion} 
In this paper, we derived the second-order positive Burgers' equation using the standard Burgers hierarchy recursion operator and investigated its integrability structure. We analytically demonstrated that the resulting nonlinear partial differential equation inherits complete integrability from the classical Burgers' equation, being perfectly linearizable to a third-order dispersion equation via the Cole-Hopf transformation. Crucially, the mathematical architecture was expanded to establish the explicit third-order flow and universally formulate the $n$-th order equation mapping utilizing the combinatorial properties of Complete Bell Polynomials. 

We definitively proved integrability by constructing the explicit scalar Lax pair (zero-curvature representation), which enabled the derivation of both the differential auto-Bäcklund transform and the algebraic nonlinear superposition principle (Darboux transformations). Multiple analytical methodologies—including separation of variables, the Hirota bilinear technique, complete polynomial discrimination (CDSPM), and Lie symmetry reduction—were employed to map out the solution space, yielding explicit rational, multi-kink, and self-similar Airy wave profiles. The algebraic frameworks and exact truncation phenomena presented herein underscore the mathematical richness of the Burgers hierarchy and provide a robust theoretical foundation for modeling higher-order dispersive wave dynamics. 

\section*{Acknowledgments} 
Mr Suman Pal is thankful to the University Grants Commission (UGC), India, for providing financial support under the Senior Research Fellowship program. The authors would also like to acknowledge Nanda Kanan Pal for valuable suggestions regarding the Darboux transformation.

\end{document}